\documentclass[preprint]{aastex}

\begin{document}

\title{The Far-Infrared/Radio Correlation in Nearby Abell Clusters}
\author{Neal A. Miller\altaffilmark{1,2,3}} 
\affil{National Radio Astronomy Observatory\altaffilmark{4}, P.O. Box O, \\ Socorro, New Mexico  87801}
\email{nmiller@aoc.nrao.edu}

\and

\author{Frazer N. Owen}
\affil{National Radio Astronomy Observatory\altaffilmark{4}, P.O. Box O, \\ Socorro, New Mexico 87801}
\email{fowen@aoc.nrao.edu}

\altaffiltext{1}{Visiting Astronomer, Kitt Peak National Observatory, National Optical Astronomy Observatories, which is operated by the Association of Universities for Research in Astronomy, Inc. (AURA) under cooperative agreement with the National Science Foundation.}

\altaffiltext{2}{Based in part on observations obtained with the Apache Point Observatory 3.5-meter telescope, which is owned and operated by the Astrophysical Reseach Consortium.}

\altaffiltext{3}{and New Mexico State University, Department of Astronomy, Box
30001/Dept. 4500, Las Cruces, New Mexico 88003}

\altaffiltext{4}{The National Radio Astronomy Observatory is a facility of the National Science Foundation operated under cooperative agreement by Associated Universities, Inc.}

\begin{abstract}
A comprehensive study of the effect of the cluster environment on the far-infrared (FIR)/radio correlation in nearby Abell clusters is presented. Using the cluster radio galaxy database from \citet{mill2000} and optical spectroscopy and high resolution radio images to remove AGN, we assess the FIR/radio correlation of cluster galaxies from the centers of the clusters out well past the classical Abell radius. The FIR/radio correlation is shown to hold quite well for star forming galaxies, and the FIR and radio fluxes for cluster AGN are also well correlated. In the case of AGN, the relative radio-to-FIR fluxes are greater and the scatter in the correlation is larger than those seen for star forming galaxies. We also find that there is a rare but statistically significant excess of star forming galaxies with enhanced radio emission in the centers of the clusters, and that the degree of this enhancement is typically a factor of two or three. The FIR/radio correlation for cluster star forming galaxies is also tested against line-of-sight velocity relative to the cluster systemic velocities, but no significant correlation is found. While the radial dependence of the FIR/radio correlation is consistent with the model wherein ram pressure increases the cluster galaxies' magnetic field strengths through compression, the velocity data do not confirm this model. Although a contribution from ram pressure can not be ruled out, the thermal pressure due to the ICM alone is an equally viable alternative. The high resolution radio images largely reject the hypothesis that the increased radio emission arises from an AGN component, strengthening the claim that the change in the correlation is caused by a change in the environment of the galaxies.
\end{abstract}

\section{Introduction}

The effect of the cluster environment on member galaxies is a well-established field of study. \citet{butc1978} studied the fraction of blue galaxies in clusters as a function of redshift, and thereby demonstrated evolution in the cluster environment. In the more local universe, studies have compared cluster environments to their field counterparts. The HI content of cluster spirals has been shown to be deficient by about an order of magnitude \citep{hayn1984,caye1990}, and several studies have investigated star formation rates in clusters \citep{kenn1983,kenn1984,gava1991,moss2000}.

A natural extension of the studies on the effect of the cluster environment is that of the far-infrared (FIR)/radio correlation. Though identified much earlier \citep{vand1973}, the FIR/radio correlation was most solidly demonstrated with data from IRAS. Using the IRAS 60$\mu$m and 100$\mu$m flux densities to estimate the total emission between about 40$\mu$m and 120$\mu$m, the FIR flux of normal star forming galaxies is strongly correlated with their radio flux (most frequently measured at 20cm). The postulated cause of this correlation is the same underlying stellar population giving rise to two very different emission mechanisms \citep[for a review see][]{cond1992}. It is believed that the FIR emission is thermal and originates in dusty HII regions heated by massive stars. The radio emission is mainly non-thermal (synchrotron) but presumably arises as the same massive stars accelerate cosmic ray electrons in supernovae. If the cluster environment were to affect these two emission mechanisms in differing ways, it would be noticeable in a deviation from the standard FIR/radio correlation.

Several studies have investigated this or related effects. Some of the earlier work was done by \citet{gava1986}, who investigated the 1.4GHz (20 cm) radio emission of spiral galaxies in Abell 1367 and Abell 1656. They found that the radio emission of the cluster spirals was up to a factor of ten times greater than that observed in comparable field spirals. Noting that some of the galaxies had high H$\alpha$ equivalent widths, they suggested that interaction of cluster spirals with the intracluster medium (ICM) enhanced star formation. Both H$\alpha$ and FIR studies have largely refuted that cluster spirals exhibit enhanced star formation \citep[e.g.,][]{kenn1984, bica1987}, requiring some other explanation for the increased radio emission relative to field spirals. \citet{gava1991} suggested that while the star formation rates of the cluster galaxies may not be greatly effected, compression of the galaxies' star forming gas could result from ram pressure with the ICM. Assuming pressure balance, the interstellar media of the cluster galaxies would reach $p_{ram}~=~\rho_{ICM} v^2$. This would increase the magnetic field strength and consequently the emitted radio power for the same star formation rate. Thus, tests of the ram pressure model focus on position in the clusters (as a proxy for ICM density) and velocity of the galaxies. In addition to position, cluster richness can serve as an indicator of density as richer cluster presumably have more significant ICMs.

\citet{ande1995} compared the FIR/radio correlation in field, rich cluster, and poor cluster environments. As the ram pressure model would predict, the rich clusters had radio over-luminous spirals (i.e., lower FIR/radio ratios) while there was no statistical difference between poor clusters and the field. Another piece of evidence for the ram pressure compression model was noted in \citet{gava1999}, which again argued that the spirals in richer nearby clusters (Abell 1367 and Abell 1656) were more radio luminous, by about a factor of five. Furthermore, they claimed that the radio over-luminous galaxies had higher velocities relative to the cluster systemic velocities, which would constitute further evidence for the ram pressure compression model. 

In this paper we investigate the FIR/radio correlation in nearby Abell clusters using the recently completed radio galaxy catalog of \citet{mill2000}. This catalog provides a number of advantages for studies of this nature. Its large angular extent within each cluster allows radial trends in the FIR/radio correlation to be explored. If enhancement of radio emission is a consequence of interaction with the ICM, it should be more prevalent in centrally-located cluster galaxies. Furthermore, the large extent within each cluster provides a built-in field sample for comparison. This removes any luminosity bias which could result in studies with field samples located nearer than the cluster samples \citep[e.g.,][]{ande1995}. The velocity data in the catalog also minimize the effect of non-cluster galaxies seen in projection on the cluster, and allow testing of the radio properties as a function of line-of-sight velocity relative to the cluster systemic velocity. And finally, since the catalog includes 329 radio galaxies spread over 18 nearby clusters, it provides a large database from which more statistically-significant conclusions can be drawn.

In addition to the advantages of the \citet{mill2000} catalog, additional observational data help solidify this project. We have reasonably high signal-to-noise optical spectra of roughly half of the radio galaxies, including most of the anamolous objects (i.e., those with unusually high radio emission relative to their FIR emission). These spectra are used to identify AGN in the radio galaxy sample. The fact that ellipticals tend to reside in more crowded environments \citep[an observation known as the morphology-density relation;][]{dres1980} and that such galaxies are more likely to host radio AGN means such identification is critical. Inclusion of these galaxies would produce a strong signal that centrally-located galaxies are radio over-luminous, but this would clearly be unrelated to any effect of environment on star formation. Since the physical scales probed by the optical spectroscopy could include an AGN as well as circumnuclear star formation, we have performed high resolution radio imaging of a subset of the sample. These radio data help constrain the net effect an AGN could have on the total radio emission from the galaxies.

The paper is organized as follows. In Section 2, we outline the data used. This includes a brief summary of the \citet{mill2000} catalog and parameters for testing, plus optical spectroscopy and high resolution radio imaging methods. The statistical tests and their results are presented in Section 3, for the complete sample and defined subsamples. Section 4 discusses the results, including the implications for the ram pressure compression model and the potential sources of bias in this investigation. Summary conclusions are provided in Section 5. For all calculations which require them, we use $H_o$ = 75 km s$^{-1}$ and $q_o~=~0.1$.

\section{Data}

\subsection{The Sample and Test Parameters}

The sample is drawn from the nearby cluster radio galaxy catalog of \citet{mill2000}. This catalog was created using the NRAO VLA Sky Survey \citep[NVSS;][]{cond1998} and covers eighteen nearby Abell clusters \citep[ACO;][]{abel1989}. Advantages of the catalog are its large physical extent within each cluster -- from the cluster cores to projected radial distances of 3$h_{75}^{-1}$Mpc -- and spectroscopic redshift confirmation of cluster members. The catalog also includes FIR flux densities from IRAS, taken either from the IRAS catalogs or determined through cross-scan addition. These flux densities, coupled with the NVSS 1.4GHz radio fluxes, are used to evaluate $q$, a parametrization of the FIR-radio correlation defined by \citet{helo1985} as

\begin{equation}
q~\equiv~\log \left( \frac{FIR}{3.75\times 10^{12}\mbox{W}~\mbox{m}^{-2}} \right) - \log \left( \frac{S_{1.4GHz}}{\mbox{W}~\mbox{m}^{-2}~\mbox{Hz}^{-1}} \right)
\end{equation}

\noindent
where FIR is defined as

\begin{equation}
\left( \frac{\mbox{FIR}}{\mbox{W}~\mbox{m}^{-2}} \right) ~\equiv~ 1.26 \times 10^{-14} \left( \frac{2.58 S_{60\mu \mbox{m}}+S_{100\mu \mbox{m}}}{\mbox{Jy}} \right) .
\end{equation}

\noindent
Star forming galaxies - both normal and starburst - are tightly grouped around $q=2.3$, having $\sigma\sim0.2$.

To assess the importance of velocity in the radio emission of the cluster galaxies, we have used the velocity data in \citet{mill2000} and \citet{stru1999} to calculate the line-of-sight deviation of each galaxy's velocity from its cluster velocity. This is simply expressed as 

\begin{equation}
\delta v~=~\frac{c| z_{galaxy} ~ - ~ z_{cluster}|}{\sigma _{cluster}} .
\end{equation}

\subsection{Optical Spectroscopy: Star Formation or AGN?}

One possible explanation for cluster galaxies with low $q$ values is that there is a contribution to the net radio flux from an AGN. While this would be an interesting result in its own right, we wish to test whether interaction with the ICM has increased the cluster spirals' radio emission via some alteration in star formation environment. Consequently, we need to identify AGN and separate them from the star forming galaxies for subsequent analysis. This was done with optical spectroscopy whenever possible, using the line diagnostics of \citet{vill1987}. Specifically, we adopt the basic definitions of \citet{ho1996}: star forming ([OI] $\lambda6300~<~0.08$ H$\alpha$, [NII] $\lambda6584~<~0.6~$ H$\alpha$, [SII] $\lambda\lambda6717+6731~<~0.4$ H$\alpha$) and AGN ([OI] $\lambda6300~\geq~0.08$ H$\alpha$, [NII] $\lambda6584~\geq~0.6~$ H$\alpha$, [SII] $\lambda\lambda6717+6731~\geq~0.4$ H$\alpha$). Note that our star forming galaxies correspond to Ho's HII nuclei, and our AGN correspond to both Seyferts and LINERs in Ho's classification. We have also included galaxies with the absorption-line spectrum indicative of an old stellar population in our AGN class. 

The spectroscopy was carried out with medium resolution gratings at both Kitt Peak and Apache Point. Basic details of the spectroscopy and two-dimensional reductions may be found in \citet{mill2000}. Classification of galaxies as star forming or AGN are presented in \citet{mill2001}, although a brief summary is provided here. For each galaxy, a 2\arcsec{} nuclear aperture and a 15\arcsec{} aperture were extracted. These enabled classification of each galaxy based on its nucleus and an off-nuclear region (the difference of the two apertures). Corrections for stellar Balmer absorption were made based on the strength of the 4000$\mbox{\AA}$ break \citep[$D_{4000}$, the ratio of average $F_\nu$ evaluated in the ranges 4050$\mbox{\AA}$ -- 4250$\mbox{\AA}$ to 3750$\mbox{\AA}$ -- 3950$\mbox{\AA}$;][]{bruz1983} using the model results of \citet{pogg1997}. Line ratio diagnostics using the corrected values are then analyzed, with the location and significance of each ratio relative to its corresponding dividing line between star forming and AGN determined. Star forming galaxies are those for which no single test produces an AGN result. This scheme is thereby conservative in that the galaxies identified as star forming are unambiguously classified as such.

The target galaxies were chosen to satisfy several science drivers, and therefore cannot be easily categorized as some subset of the complete sample. In general, the radio galaxies were observed if they either lacked public velocity measurements or had $q$ values which did not allow for simple classification as star forming or AGN (in practice, $1.0 \lesssim q \lesssim 2.1$). This has resulted in observation of all but fourteen galaxies for which $1.0 < q < 2.0$, eight of which are classified as E or S0 in NED. Galaxies with values of $q$ less than 1.0 are generally the powerful radio galaxies residing in elliptical hosts. For all observations, the nucleus of the target galaxy is placed in the slit. 

This spectroscopic selection introduces a potential source of bias. It is expected that optical spectroscopy of all the sample radio galaxies would greatly increase the number of objects in the confirmed star forming sample, and slightly increase the number in the AGN sample. This is because in general, galaxies with normal values of $q$ were not observed and these are usually normal star forming galaxies. The net effect of this on the investigation of the FIR/radio correlation in cluster star forming galaxies is somewhat uncertain. 
Inclusion of the those star forming galaxies for which long-slit spectra were not obtained could improve the significance of the results through stronger number statistics. However, these unobserved galaxies will tend to have $q\sim2.3$ independent of location within the clusters and their inclusion might in effect mask the postulated centrally-located low $q$ population. 

The spectroscopic selection also may effect the results for the AGN. Since galaxies with $q<1.00$ were assumed to be powerful radio galaxies and few optical spectra were collected for these objects, the subset corresponding to AGN is not complete. In effect, this enhances the apparent correlation of FIR and radio fluxes for the AGN sample by removing those AGN with very high radio fluxes relative to their FIR fluxes. However, the AGN sample should include almost all of the Seyferts and LINERs in the complete sample.

\subsection{High Resolution Radio Imaging}
While the optical spectroscopy can identify AGN components of the cluster radio galaxies, it suffers two drawbacks for the current study. First, the $2\arcsec$ extraction aperture corresponds to a linear resolution of about one kpc for the average cluster galaxy. The scale of a circumnuclear starburst can be significantly less than this figure, so if the emission line contribution due to an AGN were relatively weak compared to a concurrent nuclear starburst the AGN would, in effect, be missed. As such AGN could explain the enhancement of radio luminosity implied in the low $q$ galaxies, another test is necessary. Second, should AGN be present in low $q$ radio galaxies it would be useful to place limits on their net contribution to the radio luminosities. 

To achieve these goals, we have conducted a program of high resolution radio imaging on a sample of radio galaxies in Abell 1367 and Abell 1656. Using the VLA in its A configuration at a frequency of 8.46GHz, we obtain a resolution of roughly $0.25\arcsec$ or approximately 100 pc. In general, the observational goals and experimental design were the same as those used by \citet{cond1991} in their high-resolution radio survey of ultraluminous IR galaxies. The resolution of the observations implies that the radio contribution from a compact AGN would be unresolved. \citet{cond1992} finds that the most luminous starbursts have compact radio emission with $D~\sim~200$ pc, so in general these radio observations would resolve radio emission from star formation in all but the most compact nuclear starbursts.

The observations were performed during a 12-hour track of time in June 1999. Even at frequencies as low as 8.46GHz, temporal changes in atmospheric water vapor can lead to significant phase variations on the longer baselines. This is particularly true during summer days. Consequently, the observations were performed using the technique of fast switching \citep[see discussion in][]{cari1996} which relies on `nodding' the VLA dishes between the target field and a nearby phase calibrator on time scales short enough to calibrate out the atmospheric phase variations. We adopted a net cycle time of 100 seconds, with 70 seconds devoted to the target field and 30 seconds devoted to the phase calibrator, which was typically within five degrees of the target field. Note that these times include move time for the dishes, so the net time on source is around 20 seconds less per cycle. Flux calibration was performed via observation of 3C286.

A total of 20 radio galaxies -- 10 from each cluster -- were observed. The primary science targets were identified radio galaxies with low $q$ values, though additional comparison sources were also observed. These comparison sources included several normal star forming galaxies with $q$ values near the canonical value of 2.3, and several known AGN with low $q$ values. The results will be presented in the next section.

\section{Analysis}

\subsection{Statistical Results}

\begin{figure}
\figurenum{1}\label{fig1}
\epsscale{0.8}
\plotone{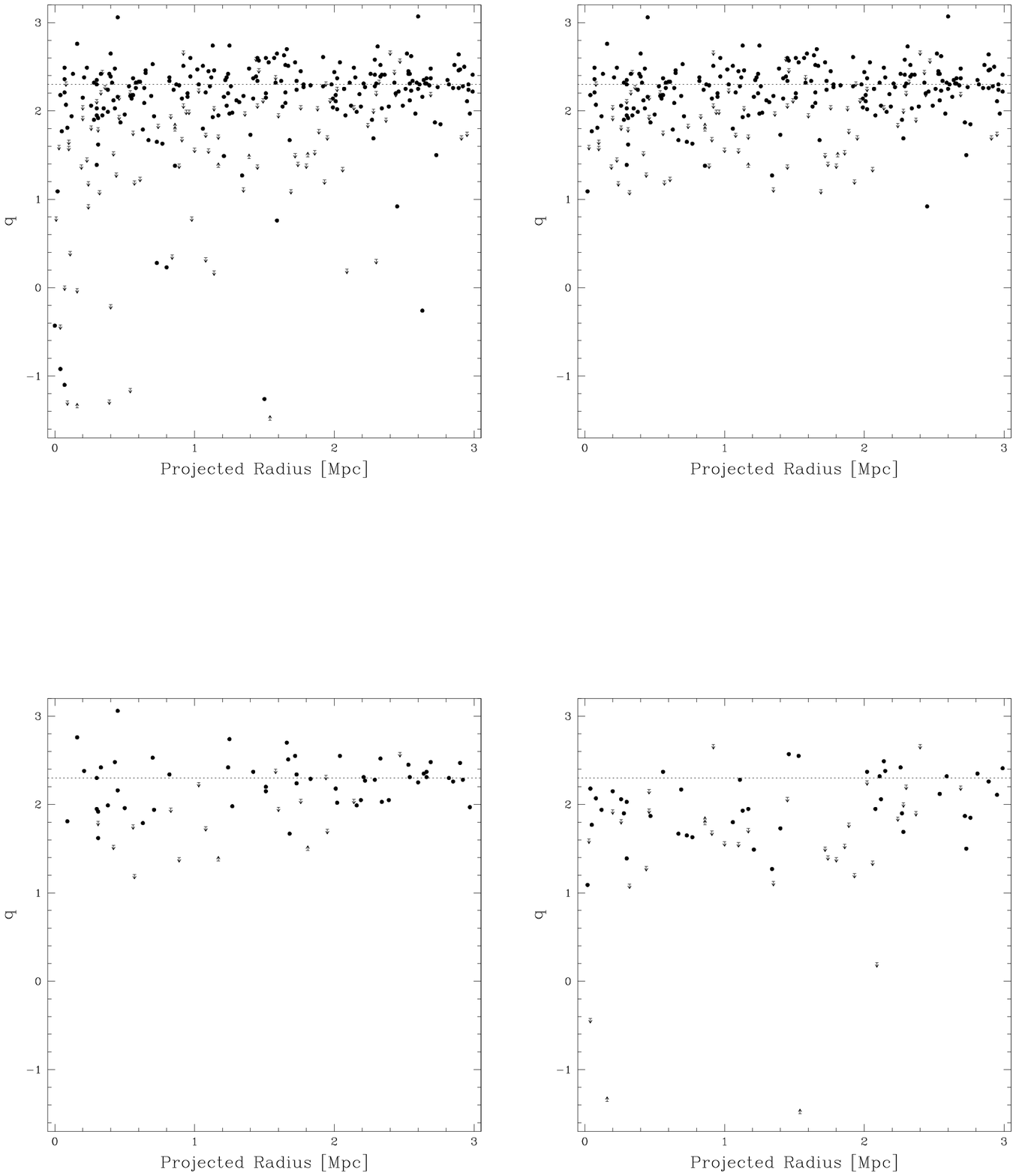}
\caption{Radial plots of $q$ values. Filled circles represent $q$ values for which the radio source was detected at both $60\mu m$ and $100\mu m$, and arrows denote upper limits where the radio source was not detected in either or both of these bands. The upper limits are calculated assuming a value of $3\times$ the rms noise in the non-detected band(s). Arrows for lower limits denote cases where the radio emission was blended among multiple objects and is not all associated with the identified galaxy. The dotted line represents the standard FIR/radio correlation. (a) Top left - the complete sample; (b) Top right - the complete sample for which $log(P_{1.4GHz}) \leq 22.7$; (c) Bottom left - spectroscopically-confirmed star forming galaxies; (d) Bottom right - spectroscopically-confirmed AGN.}
\end{figure}

As can be seen in Figure \ref{fig1}a, the FIR/radio correlation holds very well in the nearby Abell clusters of the sample. The $q$ values are mainly clustered around the expected value of 2.3 \citep[see ][]{cond1992}. Perhaps the most remarkable aspect of this is that the full sample contains a large fraction of AGN and powerful radio galaxies. The results specific to various categories of galaxies will be discussed below.

\placefigure{fig1}

The FIR/radio correlation, as parametrized by $q$, was tested for linear correlation with both radial separation from the cluster core and relative velocity. The purpose of these tests was to assess whether $q$ values of galaxies were a function of location or velocity within clusters. Because the area surveyed in each cluster reaches from the core out to large radii more consistent with the field, a significant correlation between $q$ and $r$ would suggest that the cluster environment did effect the FIR/radio correlation. Conversely, no correlation would suggest that any deviations in $q$ were normal fluctuations not caused by environment. In the case of velocity, we wish to test whether higher velocity objects are more likely to have enhanced radio emission, as would be expected in a simple model wherein ram pressure compressed the galaxies' magnetic fields. 

\begin{figure}
\figurenum{2}\label{fig2}
\plotone{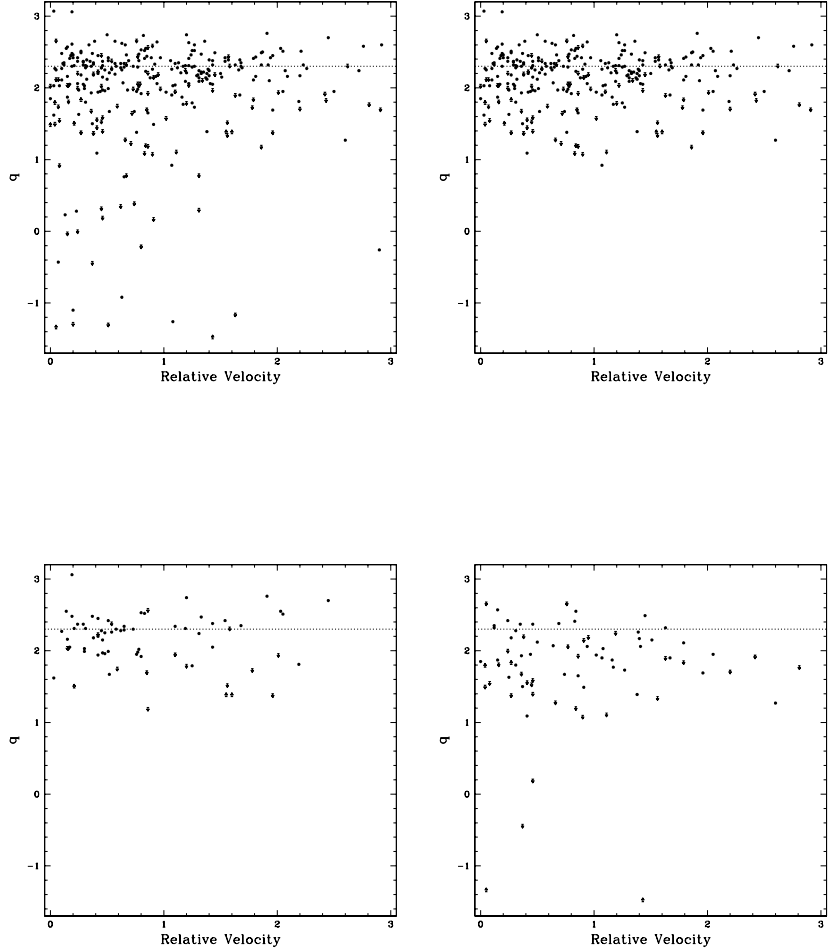}
\caption{Plots of velocity vs. $q$ values. Velocity is defined in absolute terms relative to the cluster systemic velocity, normalized by the cluster velocity dispersion. The symbols are the same as those in Figure \ref{fig1}. (a) Top left - the complete sample; (b) Top right - the complete sample for which $log(P_{1.4GHz}) \leq 22.7$; (c) Bottom left - spectroscopically-confirmed star forming galaxies; (d) Bottom right - spectroscopically-confirmed AGN.}
\end{figure}

The statistical test used was Kendall's Tau, with censoring of the $q$ values when they represented upper limits. This test examines the null hypothesis that no correlation is present between the two variables being tested. For this and all subsequent tests the IRAF `statistics' package, which implements the ``Astronomical Survival Analysis" programs \citep[see][]{feig1985,lava1992}, was used. Upper limits on $q$ values resulted from cases where the galaxy was not detected at 60$\mu$m, 100$\mu$m, or both and a value of three times the rms noise was substituted for the undetected flux(es). In a few cases, lower limits for $q$ were applied. These resulted when the low resolution of the NVSS blended the radio emission ascribed to one galaxy with another. The tests for linear correlations were performed on the sample as a whole and subsamples designed to isolate the star forming galaxies. Since we only possess optical spectroscopy for about half of the sample, we also created a subsample on the basis of radio powers alone. Based on the local RLF of \citet{cond1988}, star forming galaxies are more common than AGN for $\log(P_{1.4GHz})\leq22.7$, so we have used this figure as a rough separator of star forming galaxies and AGN. Note that this subsample does contain AGN, many of them identified in the optical spectroscopy \citep[for more discussion on this point, see][]{mill2001}. However, it provides the benefits of a simple, objective classification for galaxies and a large sample size for statistical testing. Table \ref{tbl-1} presents the results for each of the four samples (the complete sample, the subsample for $\log(P_{1.4GHz})\leq22.7$, the spectroscopically-confirmed star forming galaxies, and the spectroscopically-confirmed AGN). Graphical representations of the data are provided in Figure \ref{fig1} and Figure \ref{fig2} for $q$ vs. $r$ and $q$ vs. $\delta v$, respectively. It should also be noted that the FIR and radio fluxes for the AGN are also well correlated. As can be seen in Figure \ref{fig1}, the scatter in this correlation is larger than it is for the star forming galaxies and the radio fluxes are in excess of those predicted by the FIR/radio correlation for normal star forming galaxies. This will be discussed further in Section 4.

\placetable{tbl-1}
\placefigure{fig2}

These linear correlation tests produced evidence that radio over-luminous galaxies were centrally located in clusters, but not that they were higher velocity objects. For the complete sample (Figure \ref{fig1}a), the correlation of $q$ and $r$ was highly significant. This is easily explained as the complete sample consists of all radio galaxies in the clusters, including the powerful radio galaxies. These are almost always ellipticals, so the strength of correlation is simply a consequence of the morphology-density relation. The correlation is also quite strong for the subsample defined by radio power (Figure \ref{fig1}b). Most interestingly, the spectroscopically-confirmed star forming galaxies show a possible correlation of $q$ with $r$ ($91.2\%$; see Figure \ref{fig1}c). This is the result to be expected if the ICM affects these galaxies, and is explored in greater detail in subsequent tests. The equivalent plots for $q$ and velocity are shown in Figure \ref{fig2}a--\ref{fig2}d. None of the tested samples showed significant evidence for a correlation of $q$ with velocity (see Table \ref{tbl-1}).

\begin{figure}
\figurenum{3}\label{fig3}
\plotone{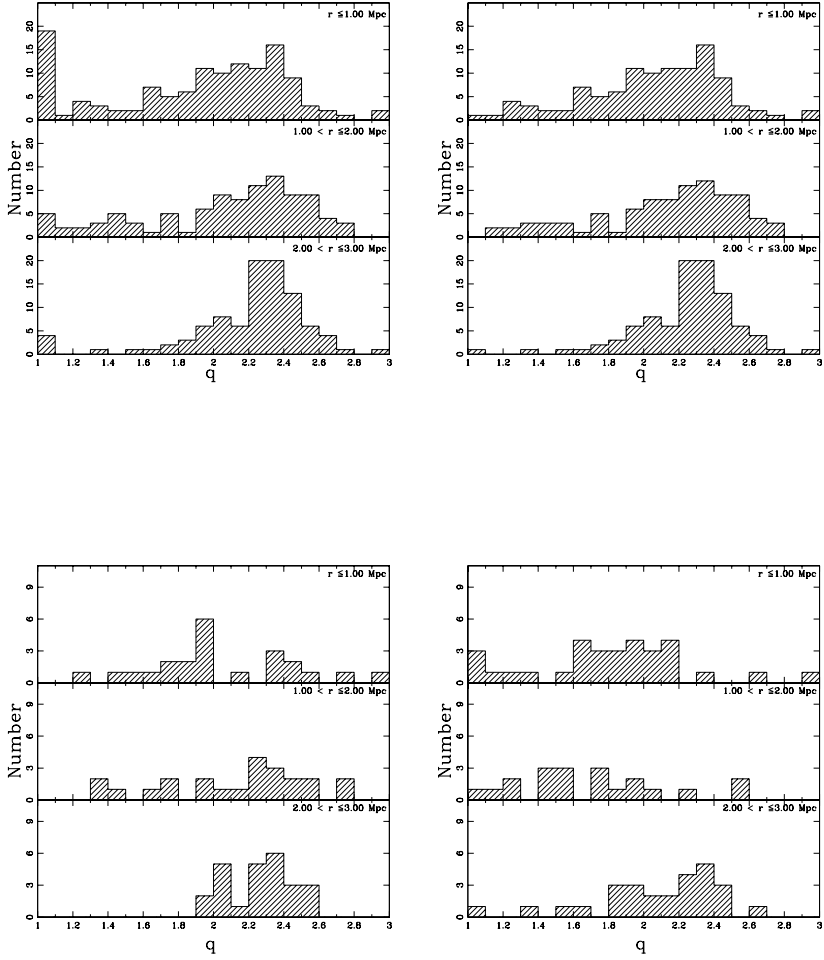}
\caption{Distributions of $q$ values, broken down by projected radial separation from the cluster cores. $q$ values less than 1.00 or greater than 3.00 are placed in the first and last bin, respectively. For the purpose of these displays, upper limits are considered detections but are weighted appropriately for the statistical tests. (a) Top left - the complete sample; (b) Top right - the complete sample for which $log(P_{1.4GHz}) \leq 22.7$; (c) Bottom left - spectroscopically-confirmed star forming galaxies; (d) Bottom right - spectroscopically-confirmed AGN.}
\end{figure}

Given the large radial extent of the survey, a linear correlation of $q$ and $r$ could easily be washed out. This is compounded by projection effects, as we can reasonably expect the galaxies at low apparent radial separation to include a fraction of galaxies with much greater true cluster-centric separations. This population could easily overwhelm the contribution of the real cluster core members. In order to test whether the $q$ values of centrally-located cluster galaxies differed from the more field-like cluster galaxies at the periphery of the clusters, we grouped the data radially and compared the $q$ value distributions. Thus, within each sample we compared the distributions of $q$ values for galaxies within one Mpc of the cluster cores (in projection) with those located between one and two Mpc, and with those between two and three Mpc. The statistical tests used were the Kolmogorov-Smirnoff (KS) test and several variants on the Wilcoxon test. The KS test evaluates the maximum difference between the cumulative distributions of two samples, and tests the null hypothesis that the samples are drawn from the same parent distribution. The Wilcoxon test ranks the data of two samples, and tests whether the samples are drawn from two separate distributions having different means. Application of the KS test used no censoring of the data (i.e., information on upper and lower limits for $q$ was ignored) while the Wilcoxon tests applied several weighting schemes for censored data \citep[see][]{feig1985}. Table \ref{tbl-2} presents the results for these tests. For the Wilcoxon tests, three types of censoring are presented -- Gehan permutation variance, logrank, and Peto \& Prentice -- though the results are nearly identical. Figure \ref{fig3} has the histograms of the $q$ distributions for each subsample and radial bin.

\placetable{tbl-2}

The results of the distributional tests are consistent with the tests of linear correlation. Each of the three bins differed significantly from the others in the analysis of the complete sample, with the strongest difference being that between the inner and outer Mpc (at a significance level much greater than $99.9\%$; see Figure \ref{fig3}a). By the outer regions of the cluster the populations are more similar, with the middle and outer Mpc bins differing at only $\sim90\%$ significance. As noted before, this is largely a result of the centrally-concentrated radio powerful ellipticals. The galaxies of the spectroscopically-confirmed star forming subsample also show a significant difference between the inner and outer Mpc bins (Figure \ref{fig3}c). The Wilcoxon tests place the significance level at $\sim97\%$, with the mean $q$ value of the inner bin being less than that of the outer bin. Furthermore, the KS test of these two groups suggests that they differ at greater than $99\%$ significance. The distributions of the AGN also seem to show some importance for the cluster environment. The Wilcoxon tests show that the middle and outer bins are most disparate (see Figure \ref{fig3}d), the result of the middle bin having the lowest mean $q$ value and the outer bin having the highest. As can be seen from Figure \ref{fig1}d, by the outer Mpc the AGN have $q$ values much closer to those expected for normal star forming galaxies.

\placefigure{fig3}

\begin{figure}
\figurenum{4}\label{fig4}
\plotone{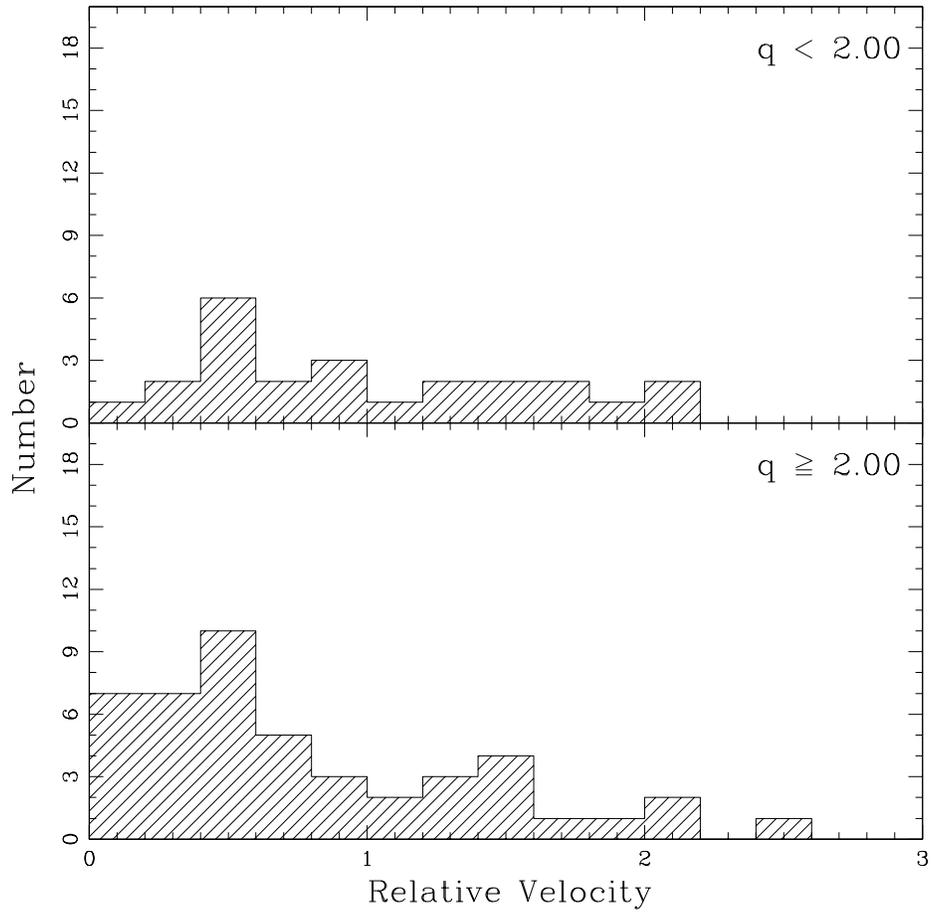}
\caption{Velocity distributions for galaxies with low $q$ ($q<2.00$; top histogram) and those with normal values of $q$ ($q \geq 2.00$; bottom histogram). Velocity is defined in absolute terms relative to the cluster systemic velocity, normalized by the cluster velocity dispersion.}
\end{figure}

The possible dependence of $q$ on velocity was also tested further. Since only the line-of-sight component of the velocity can be determined from the redshift data (analagous to the problems encountered when examining radial separations), distributional tests were performed. To test whether the low $q$ star forming galaxies have higher velocities than the normal $q$ objects, we arbitrarily divided the spectroscopically-confirmed star forming sample into those for which $q<2.00$ and those for which $q\geq2.00$; a dividing line where radio emission is a factor of two greater than the canonical value for the FIR/radio correlation. These two subsamples were then compared via the KS and Wilcoxon tests, and can be seen graphically in Figure \ref{fig4}. There is at best marginal evidence for the low $q$ galaxies to have higher relative velocities - the KS test indicates a $64\%$ probability that the two samples are drawn from different velocity distributions whereas the Wilcoxon tests are significant at about the $90\%$ level.

\placefigure{fig4}

\subsection{Results of High Resolution Radio Imaging}

The 8.46GHz observations of radio galaxies in Abell 1367 and Abell 1656 are summarized in Table \ref{tbl-2}. Only five of the twenty targets were detected, and these correspond to five AGN designated as a comparison sample. Consistent with the theory that their radio emission originates predominately from a compact active nucleus, four of these five galaxies were unresolved and the fifth consisted of a strong unresolved component within a weaker, resolved emission complex. The implied brightness temperatures of these sources also point weakly to their being AGN ($T_b~\gtrsim ~10^4~K$). The five star forming galaxies for which $q\sim2.3$ were not detected, nor were any of the ten program galaxies.

\placetable{tbl-3}

The sample selection implies that all of the objects will have radio emission at 8.46GHz, so the lack of detections are the result of the beam size and rms noise for each observation. Two important details can be obtained from this knowledge. First, an estimate of the maximum contribution to the 1.4GHz emission caused by an AGN can be determined. This is simply obtained from the rms noise per beam in the 8.46GHz map. Second, we can determine a minimum physical size over which the radio emission originates by assuming it is spread uniformly. In total, these results argue that the radio emission of the low $q$ galaxies does not arise from either an AGN or a compact nuclear starburst.

The non-detections demonstrate that the low $q$ values are not the result of radio contributions from active nuclei. Adopting a spectral index of 0.8 (i.e., $S_\nu~\propto~\nu ^{-0.8}$) we can calculate a 3$\sigma$ upper limit to the 1.4GHz contribution of an AGN. Based on the rms noise results for each observation (see Table \ref{tbl-3}), this ranges from 0.5 to 1.0 mJy. Removing this from the NVSS flux for each galaxy has a minimal effect on $q$, and even for sources at the NVSS detection limit (about 2.5 mJy) can not explain the factor of 2--3 amplification in radio fluxes implied by the lower $q$ values. For the worst case scenario of a galaxy just detected by the NVSS at 2.5 mJy, the difference in $q$ caused by 1.0 mJy of the total flux arising from a compact core is only $\sim0.2$. For the strong star forming galaxies such as 114349+195811, the difference is very nearly zero ($0.01$). Note that the assumption of a spectral index of 0.8 is probably conservative in this regard. Should the spectral index be flatter, the net contribution of the AGN to the overall 1.4GHz flux would be even less.

Interestingly, two of the test objects exhibit AGN spectra but were not detected by the 8.46GHz observations. The optical spectrum of each of these galaxies is that of a weak LINER, dominated by an old stellar population but with weak emission of [NII] and [SII]. Both are resolved by the NVSS, suggesting that the radio emission arises over significantly larger scales than those probed by the 8.46GHz observations. The radio emission could arise from a low surface brightness jet, and is also consistent with the theory in which shock ionization over fairly large scales is the emission mechanism in LINERs \citep{veil1995}.

The non-detections also place interesting limits on the size scale over which star formation is occuring in the galaxies. If the 8.46GHz emission were relatively concentrated, it would be detectable. In fact, this is the case for the one comparison sample object with a small resolved component. Assuming the 8.46GHz emission is spread uniformly, we can calculate the size at which we would just fail to detect the emission. Setting our detection threshold at 3$\sigma$, this corresponds to regions of about 400 pc diameter for the weaker sources and nearly 1.5 kpc for the strongest sources. Given the size of a compact starburst ($\sim$200 pc) and the propagation distance of the cosmic ray electrons presumed to be accelerated in supernovae within the star forming regions ($\sim$1 kpc), we can largely rule out that the star formation occurs in a compact circumnuclear starburst \citep{cond1992}.

\section{Discussion}

\subsection{Statistical Results and Implications for Models}

In all, 24 of the galaxies in the spectroscopically-confirmed star forming sample had $q<2.00$. In three of these galaxies, the $q$ value is a lower limit where the net radio flux arises from multiple blended sources. Of the remaining low $q$ galaxies, about half had $1.90\leq q <2.00$ (10 galaxies, though two represent upper limits). Thus, the population of galaxies with the enhanced radio emission implied by large deviations from the FIR/radio correlation is fairly rare. It should also be noted that the majority of the low $q$ galaxies come from richer clusters. Although seven of the low $q$ galaxies are members of Abell Richness Class 0 clusters (Abell 262, Abell 347, Abell 397, and Abell 569), the velocity dispersions of these clusters are generally around 600 km s$^{-1}$, suggesting that they are fairly massive. The cluster richest in galaxies for which $q<2.00$ is Abell 1656 (Coma), with seven such galaxies. While this hints at the importance of a richer ICM we note that Abell 426 (Perseus), a Richness Class 2 cluster like Coma, has no low $q$ galaxies. This cluster is thought to be dynamically relaxed \citep[although see][]{dupk2000} and possesses a very low fraction of spiral galaxies. 

The AGN prove interesting on two counts. First, as noted previously their FIR and radio fluxes are also strongly correlated, as has also been found in other studies \citep[e.g., studies of Seyferts by][]{baum1993,aroy1998}. This seems to be evidence for star formation in such galaxies. In fact, \citet{baum1993} were able to recover the standard FIR-radio correlation by subtracting off the radio emission from the central kpc of Seyfert galaxies, suggesting normal star formation in their disks. Second, there is evidence for a change in the FIR-radio correlation of AGN within clusters. This may be understood if some aspect of the cluster environment affects the radio emission associated with star formation, producing a radial effect in correlations for both purely star forming galaxies and for AGN with concurrent star formation.

Our spectroscopic classification of galaxies underscores a potential pitfall in studies of the radio emission from cluster spirals. Figure \ref{fig3}d and Table \ref{tbl-2} demonstrate that AGN also are affected by the cluster environment, with the $q$ values of AGN in the inner regions of the clusters being lower than those in the outer regions of the clusters. Removing the AGN which have spectra dominated by an old stellar population (i.e., creating a subsample of only emission-line AGN including both LINERs and Seyferts) does not alter this conclusion. Performing the Wilcoxon tests on the emission-line AGN within 2 Mpc (about one Abell radius) vs. those outside 2 Mpc, we find the $q$ value distributions differ at greater than $98\%$ significance. Studies which rely on imaging to select star forming galaxies are thereby prone to overestimate the effect of the cluster environment on such types. These studies frequently use wide-band filter images to select spiral galaxies, and follow up narrow-band H$\alpha$ imaging to determine which of the spirals are forming stars. The H$\alpha$ images can not differentiate between H$\alpha$ and [NII], and it is likely that some of the supposed star forming galaxies are actually AGN. Inclusion of such galaxies will produce a false signal of the effect of the cluster environment on the radio emission of cluster star forming galaxies.

The correlation and distributional tests consistently suggest the importance of location in a cluster, but not velocity. This is perhaps best exemplified in the distributional tests for the star forming galaxies (see Table \ref{tbl-2} and Figure \ref{fig3}c). Galaxies seen in projection within the inner Mpc of the clusters differ from those over 2 Mpc from the cluster center at a very high significance level. This is strong evidence that local cluster environment has an effect on star forming galaxies, especially when one considers that the effect of projection is to lessen the magnitude of such an effect. But comparing the velocities of these low $q$ galaxies with their normal counterparts does not produce significant results. This mixed evidence for the ram pressure model will be discussed below.

The total external pressure on a galaxy due to interaction with its environment is the sum of pressures due to specific effects. For simplicity, we will include only two such pressures: ram pressure and the thermal pressure due to the surrounding ICM. The sum of these pressures may then be compared to a representative ISM pressure of the galaxies to determine the relative importance for external pressure and estimate how much compression could occur. Thus,
\begin{equation}\label{eqn:Pbalance}
P_{external}~=~\rho v^2 ~+~\frac{\rho k T}{\mu m_p}
\end{equation}

We can simply evaluate where compression of the ISM could be occuring by making some assumptions. The above equation is a function of only ICM density, temperature, and the relative velocity of a galaxy to the ICM. Each of these can be estimated for every galaxy in the sample. To evaluate the local ICM density, we use the results of models based on X-ray data. \citet{brie1992} investigated the Coma cluster using ROSAT data, and found a reasonable fit to the cluster gas distribution by a King profile \citep{king1966}:

\begin{equation}\label{eqn:king}
\rho(r)~=~\frac{\rho_o}{\left[1~+~(r/r_c)^2 \right]^{3/2}}.
\end{equation}

\noindent
Using this formula, we estimate an upper limit to the local density simply by the projected radial separation from the cluster core. We adopt $\rho_o = 6.5\times10^{-27}~g/cm^{3}$ and $r_c = 0.2~Mpc$ \citep[see][]{brie1992}. In addition, the temperature is assumed to be 8 keV \citep[about the temperature of Coma; see][]{dave1993}. Given these parameters and assuming $\mu=1$ (i.e., the ICM is purely hydrogen), we can solve Equation \ref{eqn:Pbalance} for the velocity at which the external pressure equals set increments of the representative ISM pressure for the galaxies. These increased pressures would presumably compress the galaxies' ISM, thereby increasing the magnetic field strength and emitted synchrotron power.

The adopted representative ISM magnetic field strength was 5 $\mu G$. This value is typical of the regular field strength in the disks of spiral galaxies \citep{beck1996}. Although extremal values in spiral arms may reach several times this value, the non-thermal radio emission associated with star formation
 occurs as cosmic rays travel several kpc from their progenitor supernovae \citep{cond1992}. Thus, choice of the regular field strength is more appropriate. For comparison, the same models with a 10 $\mu G$ field strength are also considered. \citet{humm1995} quote this value for the regular field strength of NGC 2276 and note that it is likely higher due to interaction with the local ICM.

\begin{figure}
\figurenum{5}\label{fig5}
\plotone{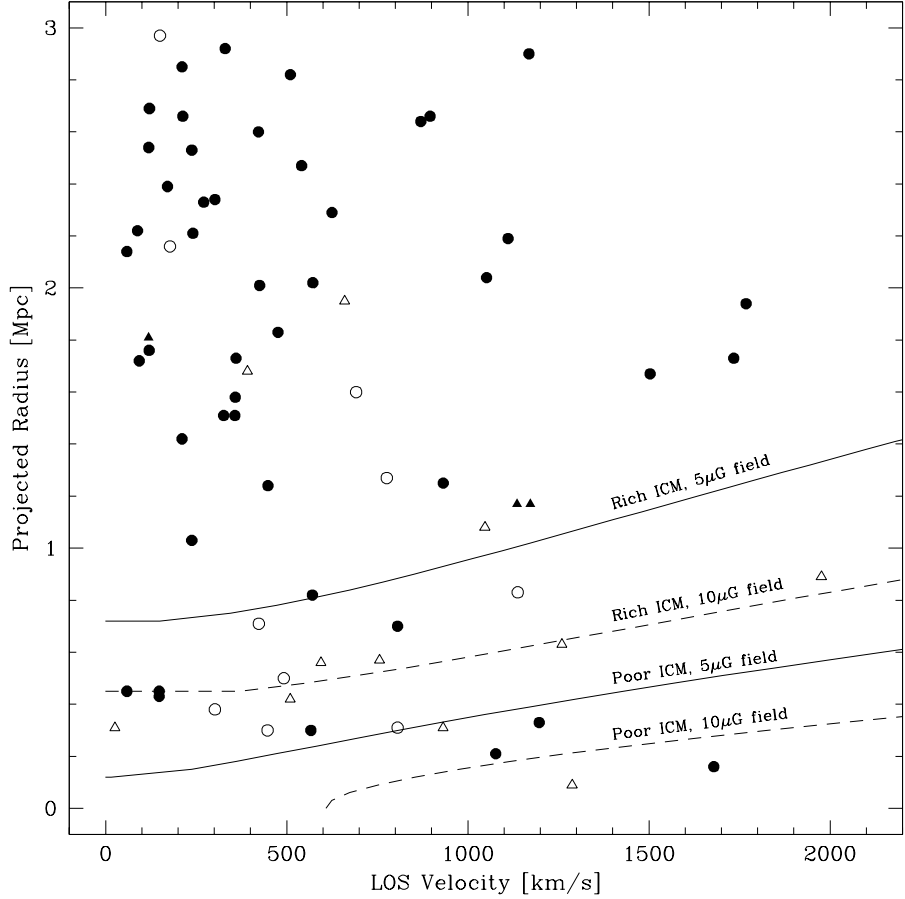}
\caption{Velocity and projected radial separation for all spectroscopically-confirmed star forming galaxies. The velocity is the absolute line-of-sight velocity relative to the cluster systemic velocity. Filled circles represent galaxies with $q\geq2.00$, open circles $1.90\leq q < 2.00$, and triangles $q<1.90$. The filled triangles are galaxies for which $q<1.90$ but the $q$ value is a lower limit. The top solid line represents the curve where the total external pressure in a dense cluster medium (ram pressure plus thermal pressure) is equal to the pressure provided by a $5\mu G$ field, the long-dashed line represents the same set of parameters but a $10\mu G$ field (see text). The bottom pair of curves are the same by for a less dense cluster medium. Deprojecting the radial component would either maintain the current ordinate or move galaxies to the right, while the net effect of velocity relative to the ICM is less clear but in general would shift galaxies up on the plot. See text for a discussion of the magnitude of the effect of projection on the radial positions.}
\end{figure}

In Figure \ref{fig5} we plot the projected radii and line-of-sight velocities for galaxies from the spectroscopically-confirmed star forming sample. To differentiate the anomalous galaxies, we code the different $q$ values by symbols. Representations of different values of external pressure are provided for several cases. The top curves assume the parameters given above for an ICM representative of the Coma cluster. The solid curve designates where the external pressure is equivalent to the magnetic pressure provided by a $5~\mu G$ field, with the dashed curve designating the same for a $10~\mu G$ field. This means that a galaxy lying along a given curve will experience twice the pressure it would experience in the absence of the cluster effects (assuming it ordinarily had a pressure consistent with the specified magnetic field strength). This would compress the galaxy's ISM, increasing the magnetic field strength, and thereby increasing the emitted synchrotron power by a factor of two. Similarly, one could interpret the 10 $\mu G$ curve as the solution to the case where a galaxy which would have a 5 $\mu G$ field in the absence of cluster effects experienced five times its non-cluster pressure. The bottom two curves adopt the same representative magnetic field strengths but assume an ICM more consistent with a poor cluster. The central density of this gas is an order of magnitude lower and the assumed temperature is 2.5 keV.

\placefigure{fig5}

The first thing to notice from Figure \ref{fig5} is that the curves are generally very flat. This provides one possible explanation for the lack of an observed correlation of $q$ with relative velocity: in general, the thermal pressure (which depends only on position within the cluster) is greater than the ram pressure. In fact, since both $P_{ram}$ and $P_{thermal}$ depend directly on $\rho$, it is easy to solve for the velocity at which they are equal. For the example of the richer cluster given above this velocity is 880 km s$^{-1}$, and for the poorer cluster it is 490 km s$^{-1}$. Only for galaxies with relative velocities above these values will ram pressure dominate over the thermal pressure. There are very few, if any, galaxies on the plot located in regions which require a significant contribution from ram pressure to the total external pressure.

This discussion is obviously highly dependent upon projection effects. To obtain an understanding of the potential magnitude of such effects, we simply apply Equation \ref{eqn:king} to the galaxy distribution in the clusters. Integrations of the equation provide the total number of galaxies within a given radius as well as the projected number of galaxies within a given radius. Inside a radius equal to the core radius, half of all galaxies seen in projection actually lie outside the core radius. Similarly, roughly a third of all galaxies seen in projection at $r\leq2r_c$ are at higher true separation, about a quarter for $r\leq3r_c$, one-fifth for $r\leq4r_c$, etc. The ratio flattens out for larger radii, meaning it is relatively constant at around $15\%$ for radii equal to or greater than $10r_c$.

The actual distribution of the star forming galaxies is somewhat uncertain. The best-fit core radius for all galaxies in the clusters is fairly small, with \citet{adam1998} citing $r_c\approx0.2$ Mpc. However, this fit is likely dominated by the strong concentration of early-type galaxies in the cluster cores. The star forming galaxies have a much larger best-fit core radius ($r_c\sim0.8$ Mpc), potentially including a `hole' in their centers \citep[e.g., see][]{morr2001,mill2001}. This argues that the statistical significance of the radial trend in $q$ values is likely much stronger than it appears.

Only in the scenario of a richer ICM do many galaxies lie in the region where external pressure is equal to or greater than the representative pressure of the galaxies' ISM. While there is a tendency for these to have lower $q$ values, a number of the low $q$ galaxies are at too large radial separations and too small relative velocities to be explained by the above model. Of course, in addition to the effect of projection noted above the true velocity of the galaxies could be higher than their line-of-sight velocities. The effect of these two biases is that galaxies will move to the right in the diagram and possibly up, which does not greatly alter the qualitative conclusions drawn. 

The observation that richer clusters have more low $q$ galaxies than poorer clusters is actually more consistent with thermal pressure being the source of compression in the galaxies than ram pressure. As can be seen in Figure \ref{fig5}, the thermal pressure provided by a poor cluster ICM is at most comparable to the ambient pressure of the galaxies. This is demonstrated by the bottom curve. The thermal pressure due to the poor cluster ICM is at most only $40\%$ of the pressure provided by a 10 $\mu G$ field, so only in regions where ram pressure is greater than thermal pressure can such a galaxy be compressed by the factor of two implied by the curve. This is the cause of the slight down turn in this curve on Figure \ref{fig5}.

Ram pressure most likely does affect some star forming galaxies, as seen from HI images of cluster spirals \citep{caye1990} and morphological studies of specific galaxies \citep[e.g., NGC 4522 and UGC6697;][]{kenn1998,nuls1982}. The simple treatment above allows a few possibilities by which ram pressure could be the important factor in explaining the low $q$ galaxies. Numerical models consistently predict that most of the galaxies infalling into clusters will be on radial orbits \citep[e.g.,][]{ghig1998}. This suggests that while few galaxies are currently seen in regions where ram pressure should be important, many galaxies will be affected by ram pressure at some point in their lives as they pass through the cluster cores. A cluster galaxy moving at 1000 km s$^{-1}$ travels roughly 1 Mpc in about $10^9$ years. These representative values for low $q$ galaxies indicate that compression due to ram pressure could occur during core passage, followed by a return to equilibrium slower than about a Gyr. 

Another possible reason for the lack of correlation of $q$ with relative velocity is that the treatment above assumed a static ICM through which galaxies moved. Bulk flows of the ICM are likely, and are frequently invoked to explain the degree of bending of extended radio sources. In fact, \citet{dupk2000} detect significant flows in Perseus through X-ray spectroscopy of iron lines performed with ASCA. Using one pointing centered on the cluster and seven additional pointings surrounding this central one, they find convincing evidence for a net rotation of the ICM. Two opposite pointings of the outer ring registered significantly different velocities for the measured line, consistent with rotation with a circular velocity of $4100~^{+2200}_{-3100}$ km s$^{-1}$ (90$\%$ confidence interval). Flows of this magnitude in a cluster previously thought to be dynamically relaxed could easily produce the lack of correlation seen for $q$ with the projected relative velocity of galaxies.

\section{Conclusions}

In this paper, we have used the cluster galaxy database of \citet{mill2000} to investigate the FIR/radio correlation in nearby Abell clusters. We find that in general the FIR/radio correlation holds remarkably well in nearby clusters of galaxies. For the star forming galaxies in the clusters we find that:

\begin{itemize}

\item Star forming galaxies with greatly enhanced radio emission are fairly rare objects. In seventeen studied clusters, only four galaxies had a radio excess of a factor of five or greater, the degree of amplification suggested by other studies. In fact, most of the anomalous objects suggested an enhancement of radio emission relative to FIR emission of only a factor of two to three.

\item There is some evidence that cluster star forming galaxies with enhanced radio emission (relative to that expected from the standard FIR/radio correlation) are centrally-located. The evidence for this effect is significant largely because of the sample size used in this study. 

\item High resolution radio images of a subset of the anomalous galaxies demonstrate that the enhanced radio emission can not be explained by a contribution from a compact AGN. In addition, the population of radio over-luminous spiral galaxies in clusters does not appear to be related to compact starbursts. 

\end{itemize}

In addition to these results, there were similar findings for cluster AGN. Although the radio powers of AGN are greater than those predicted by the canonical FIR/radio correlation value obtained for star forming galaxies, their FIR emission is also well correlated with their radio emission. As for the star forming galaxies, there is evidence that AGN located more toward the centers of the clusters have slightly greater radio emission. This argues that a portion of the net radio emission from such objects is associated with star formation.

These results have a number of implications. First, it is critical when performing studies of this nature to identify and remove AGN from the sample. Since AGN usually have stronger radio emission relative to their FIR emission, their inclusion would overstate the magnitude of the effect of the cluster environment on star formation as evidenced by departures from the FIR/radio correlation. This is likely the cause of the discrepancy between the factor of amplification of radio emission in cluster spirals found in this study (2--3) vs. those suggested by prior studies (5--10). Second, no significant correlation of increased radio emission with galaxy velocity was found. This, in conjunction with the radial distribution of the galaxies, suggests that the ram pressure model alone has difficulty explaining all of the galaxies which lie off of the normal FIR/radio correlation. While a contribution from ram pressure is possible, simple thermal pressure due to the ICM is a more likely candidate for compression of the cluster galaxies' magnetic fields.

It is hoped that additional studies will shed more light on the nature of the unusual galaxies. We are currently in the process of analyzing their spectra in detail, in hopes of learning more about their star formation histories through relative strengths of Balmer emission and absorption features. It would be interesting if the radio over-luminous star forming galaxies located farther from the cluster cores had stronger post-starburst features, which might underscore the importance of evolutionary changes as galaxies undergo their first core crossing. There is also a need for a larger sample of objects to observe at high resolution in the radio to solidify our conclusions about potential radio enhancement via AGN.

\acknowledgments
We thank Min Yun for valuable discussions and comments, and an anonymous referee for identifying useful clarifications to the manuscript. N.A.M. would like to thank the NRAO predoctoral program for support of this research.

\clearpage

\clearpage

\begin{deluxetable}{l c c c c c c c}
\tablecolumns{8}
\tablecaption{Results of Kendall's Tau Tests \label{tbl-1}}
\tablewidth{0pt}
\tablehead{
\colhead{} & \colhead{} & \colhead{} & \multicolumn{2}{c}{$q$ vs. $r$} &
\colhead{} & \multicolumn{2}{c}{$q$ vs. $\delta v$} \\
\cline{4-5} \cline{7-8} \\
\colhead{Sample} & \colhead{N} & \colhead{} & \colhead{$\tau$} &
\colhead{P} & \colhead{} & \colhead{$\tau$} & \colhead{P}}
\startdata
Full & 322\tablenotemark{a} & & 0.3037 & $>99.9\%$ & & \phs0.0730 & $72.1\%$ \\
$log(P)\leq22.7$ & 291 & & 0.2339 & $\phm{>}99.9\%$ & & \phs0.0301 & $32.4\%$ \\
Star Forming & \phn71 & & 0.2447 & $\phm{>}91.2\%$ & & $-0.0161$ & \phn$8.9\%$ \\
AGN & \phn78 & & 0.1552 & $\phm{>}78.7\%$ & & $-0.0653$ & $40.0\%$ \\
\enddata

\tablenotetext{a}{The complete catalog has 329 entries. However, seven lie in Abell 1267 for which no IRAS data are available.}

\tablecomments{The probabilities listed represent the probability that a correlation {\it is} present in the data.}

\end{deluxetable}

\clearpage

\begin{deluxetable}{l l c c c c c c c c c c c}
\tablecolumns{13}
\tablecaption{Statistical Tests of $q$ by Radial Bin\label{tbl-2}}
\tabletypesize{\small}
\tablewidth{0pt}
\tablehead{
\colhead{} & \colhead{} & \multicolumn{2}{c}{KS} & \colhead{} &
\multicolumn{2}{c}{Gehan} & \colhead{} &
\multicolumn{2}{c}{logrank} & \colhead{} &
\multicolumn{2}{c}{Peto \& Prentice} \\
\cline{3-4} \cline{6-7} \cline{9-10} \cline{12-13} \\
\multicolumn{2}{c}{Sample} & \colhead{Stat} & \colhead{Sig} & 
\colhead{} & \colhead{Stat} & \colhead{Sig} & \colhead{} &
\colhead{Stat} & \colhead{Sig} & \colhead{} & \colhead{Stat} & \colhead{Sig}}
\startdata
Complete & & & & & & & & & & & & \\ 
 & Inn-Mid & 0.182 & $\phm{>}95.0\%$ & & 2.291 & $\phm{>}97.8\%$ & & 2.090 & $\phm{>}96.3\%$ & & 2.296 & $\phm{>}97.8\%$ \\
 & Inn-Out & 0.331 & $>99.9\%$ & & 4.392 & $>99.9\%$ & & 5.166 & $>99.9\%$ & & 4.498 & $>99.9\%$ \\
 & Mid-Out & 0.196 & $\phm{>}95.3\%$ & & 1.636 & $\phm{>}89.8\%$ & & 2.523 & $\phm{>}98.8\%$ & & 1.760 & $\phm{>}92.2\%$ \\
$log(P)\leq22.7$ & & & & & & & & & & & & \\
 & Inn-Mid & 0.150 & $\phm{>}78.1\%$ & & 1.751 & $\phm{>}92.0\%$ & & 1.244 & $\phm{>}78.7\%$ & & 1.679 & $\phm{>}90.7\%$ \\
 & Inn-Out & 0.291 & $>99.9\%$ & & 3.435 & $>99.9\%$ & & 3.902 & $>99.9\%$ & & 3.765 & $>99.9\%$ \\
 & Mid-Out & 0.179 & $\phm{>}89.6\%$ & & 1.194 & $\phm{>}76.7\%$ & & 2.216 & $\phm{>}97.3\%$ & & 1.326 & $\phm{>}81.5\%$ \\
Star Forming & & & & & & & & & & & & \\
 & Inn-Mid & 0.304 & $\phm{>}76.3\%$ & & 0.095 & $\phm{>}\phn7.5\%$ & & 0.093 & $\phm{>}\phn5.7\%$ & & 0.108 & $\phm{>}\phn8.6\%$ \\
 & Inn-Out & 0.529 & $\phm{>}99.8\%$ & & 2.112 & $\phm{>}96.5\%$ & & 3.379 & $>99.9\%$ & & 2.162 & $\phm{>}96.9\%$ \\
 & Mid-Out & 0.308 & $\phm{>}79.4\%$ & & 1.351 & $\phm{>}82.3\%$ & & 2.511 & $\phm{>}98.8\%$ & & 1.516 & $\phm{>}87.1\%$ \\
AGN & & & & & & & & & & & & \\ 
 & Inn-Mid & 0.306 & $\phm{>}79.6\%$ & & 1.058 & $\phm{>}71.0\%$ & & 1.386 & $\phm{>}83.4\%$ & & 1.177 & $\phm{>}76.1\%$ \\
 & Inn-Out & 0.385 & $\phm{>}97.2\%$ & & 2.235 & $\phm{>}97.5\%$ & & 1.980 & $\phm{>}95.2\%$ & & 2.186 & $\phm{>}97.1\%$ \\
 & Mid-Out & 0.552 & $\phm{>}99.8\%$ & & 2.319 & $\phm{>}98.0\%$ & & 2.631 & $\phm{>}99.1\%$ & & 2.426 & $\phm{>}98.5\%$ \\
\enddata

\tablecomments{For each subsample, the distributions are binned by 1 Mpc. Hence, `Inn' refers to all galaxies with projected separations less than 1 Mpc from the cluster core, `Mid' refers to galaxies between 1 and 2 Mpc, and `Out' refers to those between 2 and 3 Mpc. `Sig' represents the significance level at which the distributions differ. The significance reported for each test is the probability that the two distributions differ. The Gehan test uses permutation variance, though hypergeometric variance produced the same results to within a fraction of a percent.}
\end{deluxetable}

\clearpage

\begin{deluxetable}{l c c c c c c c c c c c c}
\tablecolumns{13}
\tablecaption{Results of High Resolution Imaging \label{tbl-3}}
\tabletypesize{\small}
\tablewidth{0pt}
\tablehead{
\colhead{} & \colhead{} & \colhead{} & \colhead{} &
\multicolumn{3}{c}{1.4GHz (NVSS)} & \colhead{} &
\multicolumn{5}{c}{8.46GHz} \\
\cline{5-7} \cline{9-13} \\
\colhead{Source} & \colhead{Description} & \colhead{$q$} &
\colhead{r} & \colhead{S} & \colhead{Maj} & \colhead{Min} &
\colhead{} & \colhead{S} & \colhead{Maj} & \colhead{Min} & 
\colhead{Beam} & \colhead{Noise}}
\startdata
113654+195815 & SF\tablenotemark{a} & 2.25 & 2.60 & 18.9 & $<$21.8 & $<$18.0 & & --- & --- & --- & 0.40$\times$0.22 & 59 \\
114225+200707 & SF\tablenotemark{a} & 2.34 & 0.82 & 13.4 & \phm{$<$}23.5 & $<$22.2 & & --- & --- & --- & 0.37$\times$0.22 & 65 \\
114256+195756 & SF & 1.21 & 0.57 & 22.5 & \phm{$<$}46.9 & $<$28.5 & & --- & --- & --- & 0.34$\times$0.22 & 65 \\
114313+200016 & SF & 1.96 & 0.50 & \phn5.5 & $<$52.1 & $<$45.6 & & --- & --- & --- & 0.31$\times$0.21 & 75 \\
114349+195811 & SF & 1.62 & 0.31 & 54.8 & \phm{$<$}40.6 & $<$19.6 & & --- & --- & --- & 0.30$\times$0.22 & 67 \\
114449+194741 & SF & 1.94 & 0.12 & 17.9 & \phm{$<$}85.2 & \phm{$<$}54.8 & & --- & --- & --- & 0.23$\times$0.22 & 54 \\
114504+195824 & AGN & 1.90 & 0.28 & 11.9 & \phm{$<$}61.1 & \phm{$<$}46.0 & & --- & --- & --- & 0.21$\times$0.21 & 46 \\
114612+202329 & AGN\tablenotemark{a} & 1.58 & 1.00 & 14.9 & \phm{$<$}22.9 & $<$29.5 & & 5.95 & $<$0.06 & $<$0.06 & 0.21$\times$0.21 & 48 \\
115101+202356 & SF\tablenotemark{a} & 2.41 & 2.36 & \phn3.3 & $<$79.8 & $<$52.3 & & --- & --- & --- & 0.21$\times$0.21 & 39 \\
115243+203752 & AGN\tablenotemark{a} & 1.79 & 3.02 & \phn5.9 & $<$37.0 & $<$33.5 & & 3.55 & $<$0.05 & $<$0.05 & 0.21$\times$0.20 & 40 \\
125254+282214 & SF\tablenotemark{a} & 2.05 & 2.39 & 29.7 & \phm{$<$}23.5 & $<$20.5 & & --- & --- & --- & 0.21$\times$0.20 & 44 \\
125758+280339 & SF & 1.79 & 0.63 & \phn5.1 & $<$53.7 & $<$32.8 & & --- & --- & --- & 0.21$\times$0.20 & 50 \\
125818+290739 & SF & 1.81 & 1.83 & \phn8.6 & $<$44.4 & $<$31.2 & & --- & --- & --- & 0.21$\times$0.20 & 48 \\
130038+280325 & SF & 1.92 & 0.31 & 23.2 & \phm{$<$}30.0 & \phm{$<$}22.0 & & --- & --- & --- & 0.21$\times$0.20 & 50 \\
130041+283109 & SF & 1.40 & 0.89 & \phn8.1 & \phm{$<$}43.8 & $<$49.8 & & --- & --- & --- & 0.21$\times$0.20 & 55 \\
130056+274725 & AGN\tablenotemark{a} & 1.87 & 0.47 & 20.3 & \phm{$<$}31.0 & $<$27.4 & & 0.73 & $<$0.12 & $<$0.12 & 0.25$\times$0.21 & 41 \\
130125+291848 & AGN\tablenotemark{a} & 2.32 & 2.11 & 39.3 & $<$21.6 & $<$15.5 & & 5.76 & $<$0.06 & $<$0.06 & 0.25$\times$0.21 & 51 \\
              &                      &      &      &      &         &         & & 4.37 & \phm{$<$}0.76 & \phm{$<$}0.51 &                  & \\
130134+290749 & AGN\tablenotemark{a} & 1.55 & 1.86 & \phn6.1 & $<$43.2 & $<$29.4 & & 2.49 & \phm{$<$}0.12 & \phm{$<$}0.07 & 0.31$\times$0.21 & 60 \\
130617+290346 & AGN & 1.87 & 2.72 & 18.4 & \phm{$<$}29.9 & \phm{$<$}23.2 & & --- & --- & --- & 0.27$\times$0.21 & 53 \\
130636+275221 & SF\tablenotemark{a} & 2.28 & 2.29 & \phn7.8 & $<$36.1 & $<$35.2 & & --- & --- & --- & 0.30$\times$0.21 & 54 \\
\enddata

\tablenotetext{a}{Comparison object -- either a known AGN with low $q$, or a known star forming galaxy with $q$ near 2.3.}

\tablecomments{The galaxy descriptions are based on their nuclear optical spectra, as described in the text. Note that for 130125+291835 the 8.46GHz emission is best fit by two components. Presumably, these correspond to an unresolved AGN and a compact nuclear starburst. The galaxy itself is part of an interacting pair.}

\end{deluxetable}


\begin{thebibliography}{dummy}

\bibitem[Abell, Corwin, \& Olwin (1989)]{abel1989} Abell, G.O., Corwin, H.G., and Olowin, R.P. 1989, \apjs, 70, 1

\bibitem[Adami et al. (1998)]{adam1998} Adami, C., Mazure, A., Katgert, P., and Biviano, A. 1998, \aap, 336, 63

\bibitem[Andersen \& Owen (1995)]{ande1995} Andersen, V., and Owen, F.N. 1995, \aj, 109, 1582

\bibitem[Baum et al. (1993)]{baum1993} Baum, S.A., O'Dea, C.P., Dallacassa, D., de Bruyn, A.G., and Pedlar, A. 1993, \apj, 419, 553

\bibitem[Beck et al. (1996)]{beck1996} Beck, R., Brandenburg, A., Moss, D., Shukurov, A., and Sokoloff, D. 1996, \araa, 34, 155

\bibitem[Bicay \& Giovanelli (1987)]{bica1987} Bicay, M.D., and Giovanelli, R. 1987, \apj, 321, 645

\bibitem[Briel et al. (1992)]{brie1992} Briel, U.G., Henry, J.P., and B\"ohringer, H. 1992, \aap, 259, L31

\bibitem[Bruzual (1983)]{bruz1983} Bruzual, A.G. 1983, \apj, 273, 105

\bibitem[Butcher \& Oemler (1978)]{butc1978} Butcher, H., and Oemler, A. 1978, \apj, 219, 18

\bibitem[Carilli, Holdaway, \& Sowinski (1996)]{cari1996} Carilli, C.L., Holdaway, M.A., and Sowinski, K.P. 1996, VLA Scientific Memo. No. 169

\bibitem[Cayette et al. (1990)]{caye1990} Cayette, V., van Gorkum, J.H., Balkowski, C., and Kotanyi, C. 1990, \aj, 100, 604

\bibitem[Condon (1988)]{cond1988} Condon, J.J. 1988, in Galactic and Extragalactic Radio Astronomy, (New York:Springer-Verlag), 641

\bibitem[Condon et al. (1991)]{cond1991} Condon, J.J., Huang, Z.-P., Yin, Q.F., and Thuan, T.X. 1991, \apj, 378, 65

\bibitem[Condon (1992)]{cond1992} Condon, J.J. 1992, \araa, 30, 575

\bibitem[Condon et al. (1998)]{cond1998} Condon, J.J., Cotton, W.D., Greisen, E.W., Yin, Q.F., Perley, R.A., Taylor, G.B., and Broderick, J.J. 1998, \aj, 115, 1693

\bibitem[David et al. (1993)]{dave1993} David, L.P., Slyz, A., Jones, C., Forman, W., Vrtilek, S.D., and Arnaud, S.D. 1993, \apj, 412, 479

\bibitem[Dressler (1980)]{dres1980} Dressler, A. 1980, \apj, 236, 351

\bibitem[Dupke \& Bregman (2000)]{dupk2000} Dupke, R.A., and Bregman, J.N. 2000, \apj, in press

\bibitem[Feigelson \& Nelson (1985)]{feig1985} Feigelson, E.D., and Nelson, P.I. 1985, \apj, 293, 192

\bibitem[Gavazzi \& Jaffe (1986)]{gava1986} Gavazzi, G., and Jaffe, W. 1986, \apj, 310, 53

\bibitem[Gavazzi et al. (1991)]{gava1991} Gavazzi, G., Boselli, A., and Kennicutt, R. 1991, \aj, 101, 1207

\bibitem[Gavazzi \& Boselli (1999)]{gava1999} Gavazzi, G., and Boselli, A. 1999, \aap, 343, 93

\bibitem[Ghigna et al. (1998)]{ghig1998} Ghigna, S., Moore, B., Governato, F., Lake, G., Quinn, T., and Stadel, J. 1998, \mnras, 300, 146

\bibitem[Haynes et al. (1984)]{hayn1984} Haynes, M.P., Giovanelli, R., and Chincarini, G.L. 1984, \araa, 22, 445

\bibitem[Helou et al. (1985)]{helo1985} Helou, G., Soifer, B.T., Rowan-Robinson, M. 1985, \apjl, 298, L7

\bibitem[Ho (1996)]{ho1996} Ho, L.C. 1996, in {\it The Physics of LINERs in View of Recent Observations}

\bibitem[Hummel \& Beck (1995)]{humm1995} Hummel, E., and Beck, R. 1995, \aap, 303, 691

\bibitem[Kennicutt \& Kent (1983)]{kenn1983} Kennicutt, R.C., and Kent, S.M. 1983, \aj, 88, 1094

\bibitem[Kennicutt et al. (1984)]{kenn1984} Kennicutt, R.C., Bothun, G.D., and Schommer, R.A. 1984, \aj, 89, 1279

\bibitem[Kenney \& Koopman (1998)]{kenn1998} Kenney, J.D.P., and Koopman, R.A. 1998, astro-ph 9812363

\bibitem[King (1966)]{king1966} King, I.R. 1966, \aj, 71, 64

\bibitem[Lavalley, Isobe, \& Feigelson (1992)]{lava1992} Lavalley, M., Isobe, T., and Feigelson, E. 1992, A.S.P. Conference Series, Vol. 25, p. 245

\bibitem[Miller \& Owen (2001a)]{mill2000} Miller, N.A., and Owen, F.N. 2001a, \apjs, accepted (see astro-ph/0101114)

\bibitem[Miller \& Owen (2001b)]{mill2001} Miller, N.A., and Owen, F.N. 2001b, in preparation

\bibitem[Morrison \& Owen (2001)]{morr2001} Morrison, G.E., and Owen, F.N. 2001, in preparation

\bibitem[Moss \& Whittle (2000)]{moss2000} Moss, C., and Whittle, M. 2000, in submission

\bibitem[Nulsen (1982)]{nuls1982} Nulsen, P.E.J. 1982, \mnras, 198, 1007

\bibitem[Oke (1974)]{oke1974} Oke, J.B. 1974, \apjs, 27, 21

\bibitem[Poggianti \& Barbaro (1997)]{pogg1997} Poggianti, B.M., and Barbaro, G. 1997, \aap, 325, 1025

\bibitem[Roy et al. (1998)]{aroy1998} Roy, A.L., Norris, R.P., Kesteven, M.J., Troup, E.R., and Reynolds, J.E. 1998, \mnras, 301, 1019

\bibitem[van der Kruit (1973)]{vand1973} van der Kruit, P.C. 1973, \aap, 29, 263

\bibitem[Veilleux \& Osterbrock (1987)]{vill1987} Veilleux, S., and Osterbrock, D.E. 1987, \apjs, 63, 295

\bibitem[Struble \& Rood (1999)]{stru1999} Struble, M.F., and Rood, H.J. 1999, \apjs, 125, 35

\bibitem[Veilleux et al. (1995)]{veil1995} Veilleus, S., Kim, D.-C., Sanders, D.B., Mazzarella, J.M., and Soifer, B.T. 1995, \apjs, 98, 171

\end{thebibliography}
\end{document}